\documentclass[10pt,twocolumn,english,twocolumn,aps,prd,longbibliography]{revtex4-1}
\usepackage{lmodern}

\usepackage[T1]{fontenc}
\usepackage[latin9]{inputenc}
\usepackage{babel}
\usepackage{float}
\usepackage{amsmath}
\usepackage{graphicx}
\usepackage[unicode=true,
 bookmarks=false,
 breaklinks=false,pdfborder={0 0 1},colorlinks=false,colorlinks=true,
  urlcolor=blue,
  linkcolor=blue,
  citecolor=blue]{hyperref}

\makeatletter
\usepackage{lmodern}
\usepackage[T1]{fontenc}
\usepackage{prettyref}
\usepackage{textgreek}
\usepackage{inputenc}
\usepackage{slashed}
\usepackage{hyperref}
\usepackage{bibentry}
\usepackage{color}

\makeatother

\begin{document}
\title{Numerical approach to the semiclassical method of radiation emission for arbitrary electron spin and photon polarization}
\author{T. N. Wistisen and A. Di Piazza}
\affiliation{Max-Planck-Institut f{\"u}r Kernphysik, Saupfercheckweg 1, D-69117, Germany}
\begin{abstract}
We show how the semiclassical formulas for radiation emission of Baier,
Katkov and Strakhovenko for arbitrary initial and final spins of the
electron and arbitrary polarization of the emitted photon can be rewritten
in a form which numerically converges quickly. We directly compare
the method in the case of a background plane wave with the result obtained by using the Volkov state solution
of the Dirac equation, and confirm that we obtain the same result. We
then investigate the interaction of a circularly polarized short laser
pulse scattering with GeV electrons and see that the finite duration
of the pulse leads to a lower transfer of circular polarization than 
that predicted by the known formulas in the monochromatic case. We also see how
the transfer of circular polarization from the laser beam to the gamma
ray beam is gradually deteriorated as the laser intensity increases, entering
the nonlinear regime. However, this is shown to be recovered if the scattered
photon beam is collimated to only allow for passage of photons emitted with angles
smaller than $1/\gamma$ with respect to the initial electron direction,
where $\gamma$ is the approximately constant Lorentz factor of the electron.
The obtained formulas also allow us to answer questions regarding
radiative polarization of the emitting particles. In this respect
we briefly discuss an application of the present approach to the case
of a bent crystal and high-energy positrons.
\end{abstract}
\maketitle

\section{Introduction}

The semiclassical formalism of Baier, Katkov and Strakhovenko allows
for the approximate determination of the spectrum of emitted photons
from an ultrarelativistic electron in a virtually arbitrary external electromagnetic
field \cite{Baier1998}. For numerical applications the formulation with a single time
integration as found in \cite{Belkacem1985,PhysRevD.90.125008} for
the spin and polarization averaged result, is most useful. In this
paper we show how the basic result of the semiclassical method with
explicit electron spin and photon polarization can also be treated
numerically in a similar fashion. We use the obtained formulas in the case 
of a background plane wave, as the Dirac equation then can be solved analytically \cite{Volkov1935}, to do a direct
comparison with the spectrum obtained using the exact solution of the
Dirac equation (Volkov states)
\cite{Volkov1935,Boca_2010,PhysRevD.97.056028}. This
is the usual approach for such processes \cite{Ritus,PhysRevD.85.101701,PhysRevA.83.032106,dinu2018single,PhysRevA.80.053403,PhysRevA.83.022101,RevModPhys.84.1177,Harv09,PhysRevLett.105.080401,PhysRevD.93.085028,PhysRevLett.106.020404,Voroshilo_2015,PhysRevA.85.062102,PhysRevA.86.052104,PhysRevA.98.023417,PhysRevA.93.052102}. We consider the case of a short circularly polarized laser pulse, and find
agreement, as expected. The advantage of the presented approach is the possibility
of calculating the radiation emission under general circumstances,
i.e. also for very complicated field configurations as one only needs the classical
trajectory in the external field, which can easily be found numerically
for a given field. The presented formulas allow to
find the polarization properties of the radiation depending on the
spin of the initial and final electron, which also allows to determine
if the electrons become polarized. The latter would occur if the
spin-flip radiation has a different yield for each of the possible
initial spin states, see e.g. \cite{RevModPhys.48.417}, i.e. a generalization
of the Sokolov-Ternov effect \cite{sokolov1966synchrotron} to fields
other than that of a permanent magnetic field \cite{PhysRevA.96.043407,PhysRevLett.122.154801}. 
We briefly demonstrate this in the case of positrons channeling in a bent germanium crystal
where one has two kinds of motion superimposed, the oscillatory channeling motion between 
the bent planes, which in the unbent case would not lead to polarization, along with the motion 
along the bending arc which leads to transverse polarization of the positrons. When
the crystal is strongly bent, i.e. close to the so-called Tsyganov radius \cite{ELISHEV1979387,tsyganov1976estimates},
the polarization as in a magnetic field is obtained, while smaller
bending radii lead to smaller degrees of polarization, which the presented
method allows to predict.

Below, $e$ indicates the positron charge, and units are used, such that the fine-structure constant $\alpha$ is given by $e^2$, whereas the relativistic metric $+---$ is employed. We will use Feynman notation to write $\slashed{a}=a_\mu \gamma^\mu$, where $a^{\mu}$ is a generic 4-vector.

\section{Semiclassical approach}

Below, we study the emission by an electron of a single photon in a given background electromagnetic field. 
The basic result of the semiclassical method of Baier et al. in its
most general form for the single-photon radiation probability is expressed as \cite{Baier1998}

\begin{equation}
dP=\frac{\alpha\omega}{(2\pi)^{2}}\left|\int_{-\infty}^{\infty}R(t)e^{ik'x}dt\right|^{2}d\Omega d\omega,\label{eq:Baierfundamental}
\end{equation}
where $x^{\mu}=\{t,\boldsymbol{x}(t)\}$ is the electron 4-position
as obtained by the Lorentz force equation in the external field, $k^{\prime\mu}=\omega'\{1,\boldsymbol{n}\}$, $\omega'=\frac{\varepsilon}{\varepsilon'}\omega$,
$\omega$ is the energy of the emitted photon, $\varepsilon'=\varepsilon-\omega$,
$\varepsilon$ the electron energy, $\boldsymbol{n}$ the direction
of emission, and

\begin{equation}
R(t)=\phi_{f}^{\dagger}\left[A(t)+i\boldsymbol{\sigma}\cdot\boldsymbol{B}(t)\right]\phi_{i}.
\end{equation}
Here, $\phi_{i}$ and $\phi_{f}$ are the spinors of the initial and
final electron state (characterized by the electron 4-momentum and the electron spin in its asymptotic rest frame), $\boldsymbol{\sigma}$ denotes the vector of the Pauli spin
matrices, and

\begin{alignat}{1}
A(t) & =C\boldsymbol{\epsilon}^{*}\cdot\boldsymbol{v}(t),
\end{alignat}

\begin{alignat}{1}
\boldsymbol{B}(t) & =\boldsymbol{\epsilon}^{*}\times\left[D_{1}\boldsymbol{v}(t)+D_{2}\boldsymbol{n}\right],
\end{alignat}
with $\boldsymbol{\epsilon}$ being the polarization vector of the emitted
photon, $\boldsymbol{v}(t)=d\boldsymbol{x}(t)/dt$ being the electron velocity, and the constants being given by

\begin{equation}
C=\frac{\varepsilon}{2\sqrt{\varepsilon\varepsilon'}}\left[\sqrt{\frac{\varepsilon'+m}{\varepsilon+m}}+\sqrt{\frac{\varepsilon+m}{\varepsilon'+m}}\right],
\end{equation}

\begin{equation}
D_{1}=\frac{\varepsilon}{2\sqrt{\varepsilon\varepsilon'}}\left[\sqrt{\frac{\varepsilon'+m}{\varepsilon+m}}-\sqrt{\frac{\varepsilon+m}{\varepsilon'+m}}\right],
\end{equation}

\begin{equation}
D_{2}=\frac{\omega}{2\sqrt{\varepsilon\varepsilon'}}\sqrt{\frac{\varepsilon+m}{\varepsilon'+m}}.
\end{equation}
To evaluate the quantity in Eq. (\ref{eq:Baierfundamental}) we
need to carry out the two time integrals $\int\boldsymbol{v}(t)e^{ik'x}dt$
and $\int e^{ik'x}dt$. However, a direct computation of these
integrals converges slowly, and integrations beyond times when the acceleration
is different from zero must be included, as explained classically in \cite{Jackson1991}.
From the relations shown in \cite{PhysRevD.90.125008}, and
which are already used there in the case without polarization and spin averaging,
it is quite easy to relate these quantities to the quantities whose integrands
are proportional to the acceleration. By doing this, we have that

\begin{equation}
\int_{-\infty}^{\infty}\boldsymbol{v}(t)e^{ik'x}dt=\frac{i}{\omega'}\left(\boldsymbol{n}J-\boldsymbol{I}\right),\label{eq:velocity}
\end{equation}

\begin{equation}
\int_{-\infty}^{\infty}e^{ik'x}dt=\frac{i}{\omega'}J,\label{eq:phase}
\end{equation}
where

\begin{equation}
\boldsymbol{I}=\int_{-\infty}^{\infty}\frac{\boldsymbol{n}\times\left[\left(\boldsymbol{n}-\boldsymbol{v}\right)\times\dot{\boldsymbol{v}}\right]}{\left(1-\boldsymbol{n}\cdot\boldsymbol{v}\right)^{2}}e^{ik'x}dt,
\end{equation}

\begin{equation}
J=\int_{-\infty}^{\infty}\frac{\boldsymbol{n}\cdot\dot{\boldsymbol{v}}}{\left(1-\boldsymbol{n}\cdot\boldsymbol{v}\right)^{2}}e^{ik'x}dt.
\end{equation}
In \cite{PhysRevD.90.125008} it is shown in detail how to
calculate the electron trajectory and the quantities $\boldsymbol{I}$ 
and $J$ numerically. In particular it is appropriate to analytically carry out the cancellations
between large terms, as in e.g. $1-\boldsymbol{n}\cdot\boldsymbol{v}$
because $\boldsymbol{n}\cdot\boldsymbol{v}$ is close to $1$ for ultrarelativistic
particles. Finally, we may write

\begin{flalign}
 & \int_{-\infty}^{\infty}R(t)e^{ik'x}dt\nonumber \\
 & =-\frac{i}{\omega'}\phi_{f}^{\dagger}\left[C\boldsymbol{\epsilon}^{*}\cdot\boldsymbol{I}\right.\nonumber \\
 & \left.+i\boldsymbol{\sigma}\cdot\left(\boldsymbol{\epsilon}^{*}\times\left[\boldsymbol{I}D_{1}-(D_{1}+D_{2})\boldsymbol{n}J\right]\right)\right]\phi_{i},
\end{flalign}
and therefore we obtain the emission probability as

\begin{align}
\frac{dP}{d\Omega} & =\frac{\alpha}{(2\pi)^{2}}\frac{\omega}{\omega'^{2}}\nonumber \\
 & \times\left|\phi_{f}^{\dagger}\left[C\boldsymbol{\epsilon}^{*}\cdot\boldsymbol{I}\right.\right.\nonumber \\
 & \left.+i\boldsymbol{\sigma}\cdot\left(\boldsymbol{\epsilon}^{*}\times\left[\boldsymbol{I}D_{1}-(D_{1}+D_{2})\boldsymbol{n}J\right]\right)\right]\phi_{i}\Big|^{2}\label{eq:Baierfinal}
\end{align}

\section{Volkov-state approach}

If the background field is a plane wave, i.e. if the 4-vector potential $A^{\mu}(\varphi)$ only depends on the phase $\varphi=k_0x$, where $k_0=(\omega_0,\boldsymbol{k}_0)$ is the 4-momentum associated with the photons of the plane wave, the corresponding Dirac equation

\begin{equation}
\left(i\slashed{\partial}+e\slashed{A}-m\right)\psi=0,
\end{equation}
can be solved analytically \cite{Volkov1935}. Below we assume that the plane wave propagates along the negative $z$ direction and we choose 4-vector potential $A^{\mu}(\varphi)$ in the Lorenz gauge where $A^0(\varphi)=A^3(\varphi)=0$. The positive-energy solution reads
\begin{equation}
\psi(x)=\frac{1}{\sqrt{2\varepsilon}}\left(1-\frac{e\slashed{k}_0\slashed{A}}{2k_0p}\right)ue^{iS},\label{eq:volkovstate}
\end{equation}
where $p$ is the asymptotic 4-momentum of the electron, (we have
set the quantization volume equal to $1$), where

\begin{equation}
S=-px+\frac{e}{k_0p}\int^{\varphi}d\varphi'\left[pA(\varphi')+\frac{e}{2}A^{2}(\varphi')\right]\label{eq:Sminus}
\end{equation}
is the classical action of the electron in the plane wave, and where $u$ is a short notation for the constant vacuum bispinor (which is characterized by the electron spin in the corresponding electron rest frame and by the electron 4-momentum $p$). The leading-order matrix element for single-photon emission is given by

\begin{equation}
S_{fi}=ie\sqrt{\frac{4\pi}{2\omega}}\int d^{4}x\bar{\psi}_{f}(x)\slashed{\epsilon}^{*}e^{ikx}\psi_{i}(x),\label{eq:matrixelement}
\end{equation}
where $\psi_{i/f}(x)$ indicates the Volkov state corresponding to the initial/final electron state, and the differential probability of emission is then

\begin{equation}
dP=\left|S_{fi}\right|^{2}\frac{d^{3}p_{f}}{(2\pi)^{3}}\frac{d^{3}k}{(2\pi)^{3}}.\label{eq:probability}
\end{equation}
In the gauge we are working, the 4-potential can be written as

\begin{equation}
A^{\mu}(\varphi)=\sum_{j=1}^{2}a_{j}^{\mu}f_{j}(\varphi),\label{eq:potential}
\end{equation}
where $a_{j}^{\mu}$ are two 4-vectors such that $a_jk_0=0$ and $a_1a_2=0$ and where $f_j(\varphi)$ are two arbitrary (physically well-behaved) functions.
By setting the arbitrary phase in the indefinite integrals in the phase of Volkov states to zero, we introduce the quantities

\begin{equation}
F_{j}(\varphi)=\int_{0}^{\varphi}f_{j}(\varphi')d\varphi',
\end{equation}

\begin{equation}
G_{j}(\varphi)=\int_{0}^{\varphi}f_{j}^{2}(\varphi')d\varphi'.
\end{equation}
Then, by inserting the expressions of Eqs. (\ref{eq:volkovstate}) and (\ref{eq:potential}) into
Eq. (\ref{eq:matrixelement}), we obtain that

\begin{flalign}
S_{fi} & =ie\sqrt{\frac{4\pi}{2\omega}}\frac{1}{\sqrt{4\varepsilon_{f}\varepsilon_{i}}}\int d^{4}x\nonumber \\
 & \bar{u}_{f}\left(\slashed{\epsilon}^{*}+\sum_{j=1}^{2}\left[B_{j}f_{j}(\varphi)+C_{j}f_{j}(\varphi)^{2}\right]\right)u_{i}\nonumber \\
 & \times e^{-i(p_{i}-p_{f}-k)x}e^{i\left(\sum_{j=1}^{2}\left[\alpha_{j}F_{j}(\varphi)+\beta_{j}G_{j}(\varphi)\right]\right)},\label{eq:Sfinal}
\end{flalign}
where we have defined

\begin{equation}
\alpha_{j}=e\left[\frac{p_{i}a_{j}}{k_0 p_{i}}-\frac{p_{f}a_{j}}{k_0 p_{f}}\right],
\end{equation}

\begin{equation}
\beta_{j}=\frac{e^{2}a_{j}^{2}}{2}\left[\frac{1}{k_0 p_{i}}-\frac{1}{k_0 p_{f}}\right],
\end{equation}
and

\begin{equation}
B_{j}=-\left[\frac{e\slashed{a}_{j}\slashed{k}_0}{2k_0 p_{f}}\slashed{\epsilon}^{*}+\slashed{\epsilon}^{*}\frac{e\slashed{k}_0 \slashed{a}_{j}}{2k_0 p_{i}}\right],
\end{equation}

\begin{flalign}
C_{j} & =\frac{e\slashed{a}_{j}\slashed{k}_0}{2k_0 p_{f}}\slashed{\epsilon}^{*}\frac{e\slashed{k}_0 \slashed{a}_{j}}{2k_0 p_{i}}\nonumber \\
 & =-\frac{e^{2}a_{j}^{2}}{2\left(k_0 p_{f}\right)\left(k_0 p_{i}\right)}\left(\epsilon^{*}k_0\right)\slashed{k}_0,
\end{flalign}
with $j=1,2$ (we have set $\epsilon^{\mu}=(0,\boldsymbol{\epsilon})$). Now, we can write the functions in Eq. (\ref{eq:Sfinal})
as a Fourier transform

\begin{flalign}
 & f_{j}^{n}(\varphi)e^{i\left(\sum_{j=1}^{2}\left[\alpha_{j}F_{j}(\varphi)+\beta_{j}G_{j}(\varphi)\right]\right)}\nonumber \\
 & =\int_{-\infty}^{\infty}A_{n,j}(s,\alpha,\beta)e^{-is\varphi}ds,
\end{flalign}
where

\begin{flalign}
 & A_{n,j}(s,\alpha,\beta)\nonumber \\
 & =\frac{1}{2\pi}\int_{-\infty}^{\infty}d\varphi f_{j}^{n}(\varphi)e^{i\left(s\varphi+\sum_{j=1}^{2}\left[\alpha_{j}F_{j}(\varphi)+\beta_{j}G_{j}(\varphi)\right]\right)},
\end{flalign}
defined for $n=0,1,2$. When $n=0$, the $j$ subscript is
superfluous and we will therefore denote this function as $A_{0}(s,\alpha,\beta)$.
This function is however problematic as it diverges but it can be regularized by
using the identity (see also \cite{PhysRevA.80.053403,PhysRevA.83.022101,PhysRevA.83.032106})

\begin{alignat}{1}
0 & =\int_{-\infty}^{\infty}e^{ih(\varphi)}h'(\varphi)d\varphi,\label{eq:hint}
\end{alignat}
where

\begin{equation}
h(\varphi)=s\varphi+\sum_{j=1}^{2}\left[\alpha_{j}F_{j}(\varphi)+\beta_{j}G_{j}(\varphi)\right].\label{eq:hfunc}
\end{equation}
In this way, we obtain

\begin{flalign}
A_{0}(s,\alpha,\beta) & =\nonumber \\
 & =\frac{1}{2\pi}\int_{-\infty}^{\infty}e^{i\left(s\varphi+\sum_{j=1}^{2}\left[\alpha_{j}F_{j}(\varphi)+\beta_{j}G_{j}(\varphi)\right]\right)}d\varphi\nonumber \\
 & =-\frac{1}{s}\sum_{j=1}^{2}\left[\alpha_{j}A_{1,j}+\beta_{j}A_{2,j}\right].
\end{flalign}
By replacing these expressions in Eq. (\ref{eq:Sfinal}), and carrying out the integration over $d^{4}x$, we can write the amplitude in the form

\begin{flalign}
S_{fi} & =ie\sqrt{\frac{4\pi}{2\omega}}\frac{1}{\sqrt{4\varepsilon_{f}\varepsilon_{i}}}\int ds(2\pi)^{4}\delta^{4}(p_{i}-p_{f}-k+sk_0)\nonumber \\
 & \times\bar{u}_{f}\left(\slashed{\epsilon}^{*}A_{0}+\sum_{j=1}^{2}\left[B_{j}A_{1,j}+C_{j}A_{2,j}\right]\right)u_{i}.
\end{flalign}
Now we can use the energy delta function to fix $s$ such that

\begin{equation}
s_{0}=\frac{\varepsilon_{f}+\omega-\varepsilon_{i}}{\omega_0},
\end{equation}
and the delta function can be transformed as $\delta(\varepsilon_{i}-\varepsilon_{f}-\omega+s\omega_0)=\frac{1}{\omega_0}\delta(s-s_{0})$:

\begin{align}
S_{fi}= & ie\sqrt{\frac{4\pi}{2\omega}}\frac{1}{\sqrt{4\varepsilon_{f}\varepsilon_{i}}}(2\pi)^{4}\frac{1}{\omega_0}\delta^{3}(\boldsymbol{p}_{i}-\boldsymbol{p}_{f}-\boldsymbol{k}+s_{0}\boldsymbol{k}_0)\nonumber \\
 & \times\bar{u}_{f}\left(\slashed{\epsilon}^{*}A_{0}+\sum_{j=1}^{2}\left[B_{j}A_{1,j}+C_{j}A_{2,j}\right]\right)u_{i}.
\end{align}
At this point we would then take the norm-square to obtain the transition
probability, however we are then faced with the problem of how to take the square
of the delta-function which has the complication that $s_{0}$ is
a function of the momenta. The correct way to do this, is to consider
instead the more realistic case of an initial wave packet $\Psi_{i}(x)=\int\psi_{\boldsymbol{p}_{i}}(x)c(\boldsymbol{p}_{i})d^{3}\boldsymbol{p}_{i}$
where $\psi_{\boldsymbol{p}_{i}}(x)$ is the Volkov solution with momentum $\boldsymbol{p}_{i}$ and unindicated fixed spin quantum number. To preserve normalization we must have
that $\int\left|c(\boldsymbol{p}_{i})\right|^{2}d^{3}\boldsymbol{p}_{i}=1/(2\pi)^{3}$.
Then the momentum delta-function can be transformed as

\begin{equation}
\delta^{3}(\boldsymbol{p}_{i}-\boldsymbol{p}_{f}-\boldsymbol{k}+s_{0}(\boldsymbol{p}_{i})\boldsymbol{k}_0)=\frac{1}{|\mathcal{J}_{i}|}\delta^{3}(\boldsymbol{p}_{i}-\boldsymbol{p}_{i,\text{sol}}),
\end{equation}
where $\mathcal{J}_{i}=\partial\boldsymbol{g}/\partial\boldsymbol{p}_{i}=I-\frac{\boldsymbol{k}_0\boldsymbol{p}_{i}^{T}}{\omega_0\varepsilon_{i}}$,
is the Jacobian matrix with $\boldsymbol{g}\left(\boldsymbol{p}_{i}\right)=\boldsymbol{p}_{i}-\boldsymbol{p}_{f}-\boldsymbol{k}+s_{0}(\boldsymbol{p}_{i})\boldsymbol{k}_0$,
$\boldsymbol{g}(\boldsymbol{p}_{i,\text{sol}})=\boldsymbol{0}$ ($\boldsymbol{k}_0\boldsymbol{p}_{i}^{T}$ indicates the dyadic product between the vectors $\boldsymbol{k}_0$ and $\boldsymbol{p}_{i}$), and
so using Sylvester's determinant theorem we obtain

\begin{equation}
|\mathcal{J}_{i}|=\text{det}\left(I-\frac{\boldsymbol{k}_0\boldsymbol{p}_{i}^{T}}{\omega_0\varepsilon_{i}}\right)=1-\frac{\boldsymbol{k}_0\cdot\boldsymbol{p}_{i}}{\omega_0\varepsilon_{i}}=\frac{k_0 p_{i}}{\omega_0\varepsilon_{i}}.
\end{equation}
Therefore we finally write the transition amplitude in the form
\begin{align}
S_{fi}= & ie\sqrt{\frac{4\pi}{2\omega}}\frac{1}{\sqrt{4\varepsilon_{f}\varepsilon_{i}}}(2\pi)^{4}\frac{\varepsilon_{i}}{k_0 p_{i}}\nonumber \\
 & \times c(\boldsymbol{p}_{f}+\boldsymbol{k}-s_{0}(\boldsymbol{p}_{i,\text{sol}},\boldsymbol{p}_{f})\boldsymbol{k}_0)\\
 & \times\bar{u}_{f}\left(\slashed{\epsilon}^{*}A_{0}+\sum_{j=1}^{2}\left[B_{j}A_{1,j}+C_{j}A_{2,j}\right]\right)u_{i}.
\end{align}
Now, in order to find the probability using Eq. (\ref{eq:probability}) we take
the norm-square of the above amplitude and, having in mind the case of a narrow wave packet \cite{PhysRevA.93.052102}, replace $\left|c(\boldsymbol{p}_{i})\right|^{2}=\delta^3(\boldsymbol{p}_{i}-\boldsymbol{p}_{i,0})/(2\pi)^{3}$.
Analogously as above, we now have a delta-function which we can evaluate by integration over $d^{3}\boldsymbol{p}_{f}$
and the transformation of the delta-function yields a factor of $\frac{\omega_0\varepsilon_{f}}{k_0 p_{f}}$.
Finally, we then obtain the differential emission probability

\begin{flalign}
dP & =\left|\bar{u}_{f}\left(\slashed{\epsilon}^{*}A_{0}+\sum_{j=1}^{2}\left[B_{j}A_{1,j}+C_{j}A_{2,j}\right]\right)u_{i}\right|^{2}\nonumber \\
 & \times\frac{e^{2}}{4}\frac{1}{\left(k_0 p_{i}\right)\left(k_0 p_{f}\right)}\omega d\omega d\Omega,
\end{flalign}
which can now be evaluated numerically. The bispinors in this expression are chosen as \cite{Bere71}

\begin{equation}
u=\sqrt{\varepsilon+m}\left(\begin{array}{c}
\phi\\
\frac{\boldsymbol{\sigma}\cdot\boldsymbol{p}}{\varepsilon+m}\phi
\end{array}\right),
\end{equation}
where $\phi$ are spinors to be chosen as an orthonormal basis
of eigenstates of $\boldsymbol{\sigma}\cdot\boldsymbol{s}$,  with
$\boldsymbol{s}$ being the direction of the otherwise arbitrary spin quantization axis in the
rest frame of the electron.

\begin{figure}[t]
\includegraphics[width=1\columnwidth]{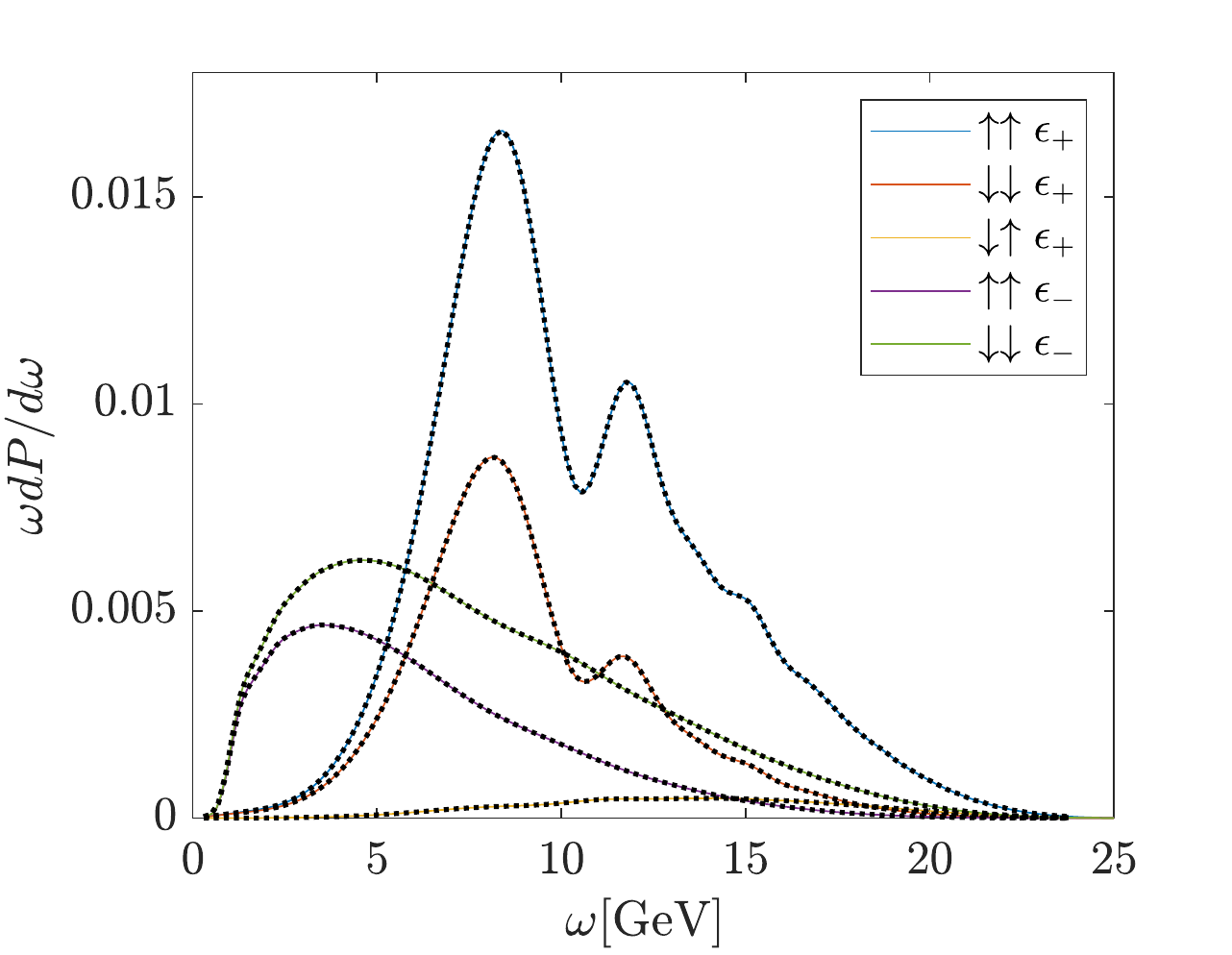}

\caption{The intensity spectrum $\omega dP/d\omega$ according to the semiclassical approach is
shown as solid lines corresponding to the different possibilities of initial and final spins and photon
polarizations. The black dotted curves on top of the solid curves indicate the same quantities but calculated using
the Volkov-states approach. We have not plotted the curve corresponding to $\uparrow\downarrow\boldsymbol{\epsilon}_{+}$
and $\downarrow\uparrow\boldsymbol{\epsilon}_{-}$ as the corresponding yields are much smaller
than the others and the curves would not be visible. Also, we have not plotted
$\uparrow\downarrow\boldsymbol{\epsilon}_{-}$ as it coincides with $\downarrow\uparrow\boldsymbol{\epsilon}_{+}$. In all these cases, the two approaches also agree.\label{fig:fig1}}
\end{figure}

\section{Discussion of results}

The above derivations were carried out without introducing a particular
plane-wave pulse. We will now consider a particular choice of the 4-vector potential
and carry out the corresponding numerical calculation using the semiclassical
method and the Volkov-states method. We set $a_{1}^{\mu}=\{0,a_{x},0,0\}$, $a_{2}^{\mu}=\{0,0,a_{y},0\}$, $k^{\mu}=\{\omega_{0},0,0,-\omega_{0}\}$, and

\begin{equation}
f_{1}(\varphi)=d(\varphi)\text{cos}(\varphi),
\end{equation}

\begin{equation}
f_{2}(\varphi)=d(\varphi)\text{sin}(\varphi),
\end{equation}

\begin{align}
d(\varphi) & =\begin{cases}
\text{sin}^{4}\left(\frac{\varphi}{2N}\right), & 0<\varphi<2\pi N,\\
0 & \text{otherwise},
\end{cases}
\end{align}
that is, we choose a pulse with envelope $d(\varphi)$ and negative
helicity (right-handed) circular polarization \footnote{Due to the presence of the finite pulse shape $d(\varphi)$ the electric field of the wave is not, rigorously speaking, circularly polarized. For the sake of simplicity, however, in the numerical examples we choose sufficiently long pulses that we can ignore this subtlety.}.
\begin{figure}[t]
\includegraphics[width=1\columnwidth]{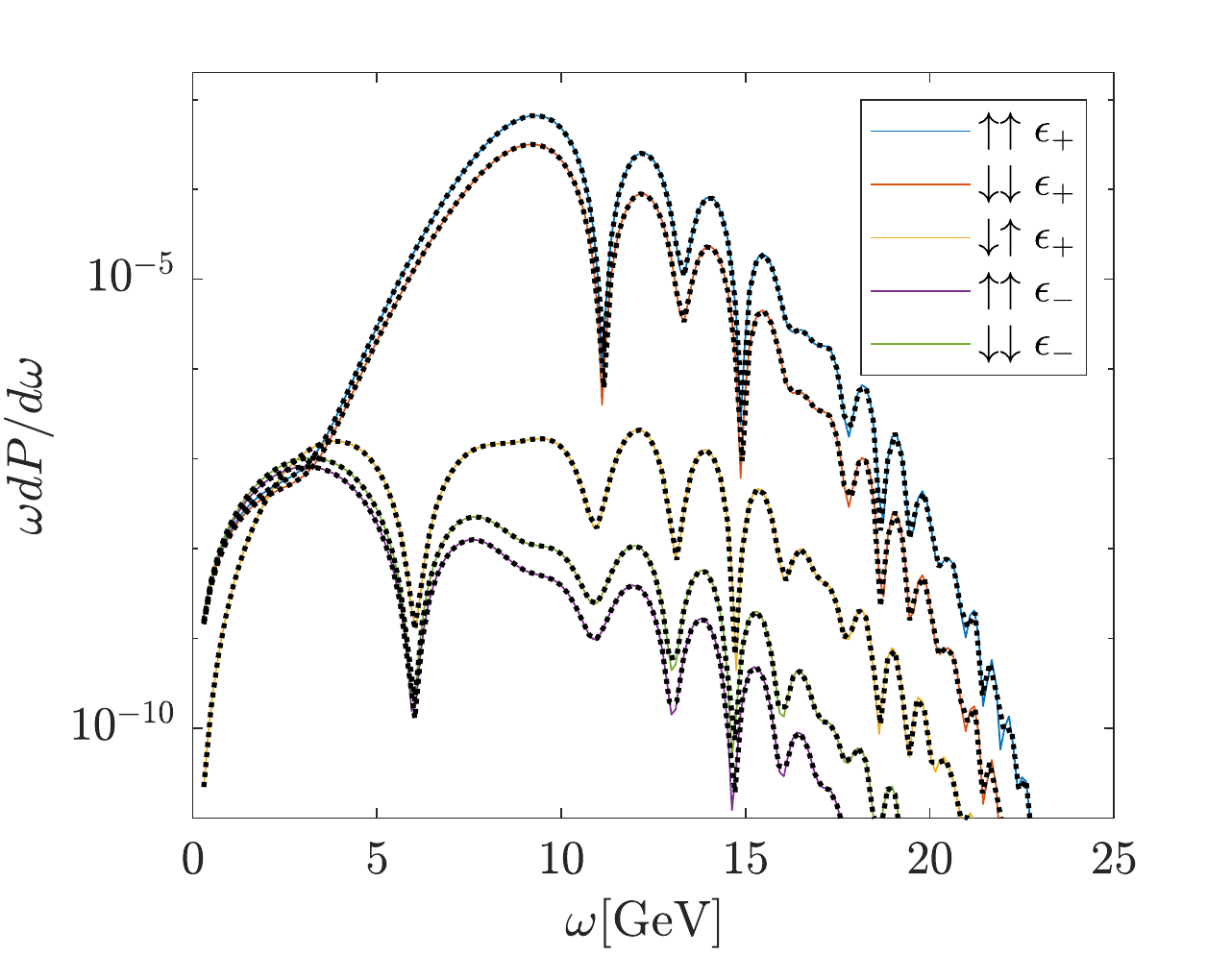}

\caption{Integrated intensity of radiation as in Fig. (\ref{fig:fig1}) but with a collimation
angle of $0.1/\gamma$, as explained in the text.\label{fig:fig2}}
\end{figure}
We define the polarizations of the outgoing light as

\begin{equation}
\boldsymbol{\epsilon}_{\pm}=\frac{1}{\sqrt{2}}\left(\boldsymbol{\epsilon}_{1}\pm i\boldsymbol{\epsilon}_{2}\right),\label{eq:eplus}
\end{equation}
where

\begin{equation}
\boldsymbol{\epsilon}_{1}=\frac{\hat{\boldsymbol{y}}\times\boldsymbol{k}}{\left|\hat{\boldsymbol{y}}\times\boldsymbol{k}\right|},\label{eq:e1}
\end{equation}
with $\hat{\boldsymbol{y}}$ being the unit vector in the $y$ direction, and

\begin{equation}
\boldsymbol{\epsilon}_{2}=\frac{\boldsymbol{k}\times\boldsymbol{\epsilon}_{1}}{\left|\boldsymbol{k}\times\boldsymbol{\epsilon}_{1}\right|}.\label{eq:e2}
\end{equation}

\begin{figure}[t]
\includegraphics[width=1\columnwidth]{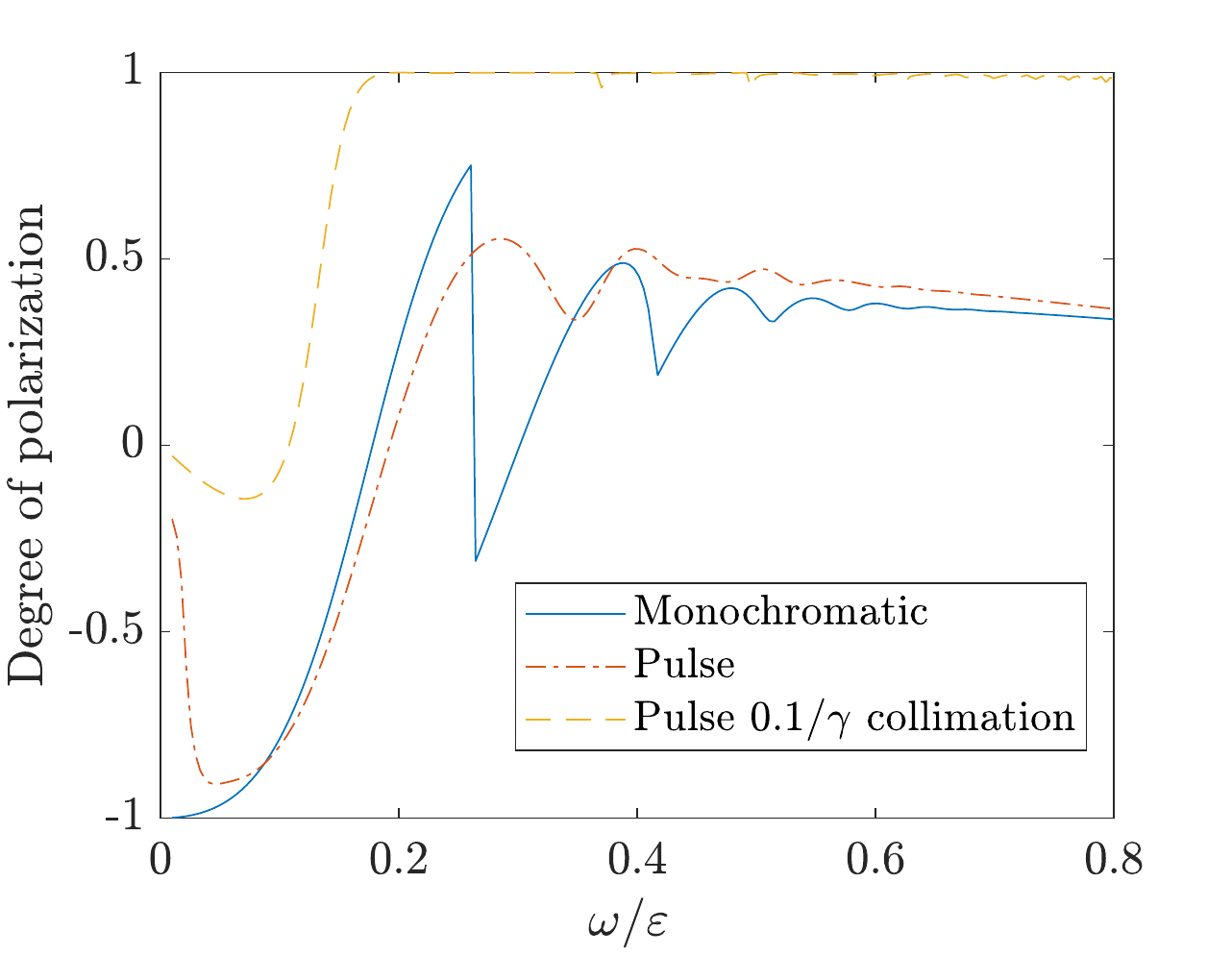}

\caption{Degree of circular polarization of the Compton scattered radiation
for nonlinear Compton scattering in a monochromatic wave without collimation,
and in the short pulse described in the text with and without angular
collimation.\label{fig:fig3}}
\end{figure}

According to this choice $\boldsymbol{\epsilon}_{1}$ and $\boldsymbol{\epsilon}_{2}$
are unit vectors orthogonal to each other and to $\boldsymbol{k}$
and such that, if $\boldsymbol{k}$ lies along the $z$ axis, they indicate the polarization along the $x$ and $y$ direction, respectively. The $\boldsymbol{\epsilon}_{\pm}$ basis corresponds to circular
polarization with helicity of $\pm1$. As the spin basis we have chosen
quantization axis along the $z$ direction such that $\phi$ may be
chosen as $\left(\begin{array}{cc}
1 & 0\end{array}\right)^{T}$or $\left(\begin{array}{cc}
0 & 1\end{array}\right)^{T}$, denoted by $\uparrow$ and $\downarrow$, respectively in the figures.
We set $a_{x}=a_{y}=m\xi/e$, where $\xi$ is the classical nonlinearity
parameter, which we have set $\xi=1$, $N=5$, and the electron energy
$\varepsilon=30$ GeV for the Figs. (\ref{fig:fig1}), (\ref{fig:fig2})
and (\ref{fig:fig3}). Since the typical emission angles are small, we write
$k_{x}=\omega\theta_{x}$ and $k_{y}=\omega\theta_{y}$ and then
$d\Omega=d\theta_{x}d\theta_{y}$. In Fig. (\ref{fig:fig1}) we
have restricted the angular integration such that $|\theta_{x}|<(\xi+3)/\gamma$,
where $\gamma$ is the initial Lorentz factor of the electron, 
and the same for $\theta_{y}$ such that nearly all emitted radiation
is included. In this figure we compare the semiclassical approach
based on the formulas of Baier, Katkov and Strakhovenko and compare
with the results obtained using the Volkov states. The results indicate
nearly perfect agreement between the two approaches, which is expected
since the motion in a plane wave is intrinsically semiclassical \cite{Ritus}. In Fig. (\ref{fig:fig2}) we do the same but
restrict the emission angles over a smaller interval (collimation)
i.e. $|\theta_{x}|<0.1/\gamma$ and the same for $\theta_{y}$. In
this case the emitted radiation with negative helicity is highly suppressed
and therefore we plotted the results on a logarithmic scale. This
is expected due to angular momentum conservation along the $z$ axis. Since the electron flipping its spin is unlikely for ultrarelativistic electrons \cite{Bere71}, the outgoing
light must have opposite helicity as that of the laser field to conserve
angular momentum. Finally, the agreement between the semiclassical
method and the Volkov-state method in this case indicates an agreement
of the two approaches also at the level of angularly resolved spectra.

In Fig. (\ref{fig:fig3}) we show how the collimation affects the degree of circular polarization, defined as

\begin{equation}
\mathcal{P}=\frac{\frac{dP^{+}}{d\omega}-\frac{dP^{-}}{d\omega}}{\frac{dP^{+}}{d\omega}+\frac{dP^{-}}{d\omega}}.\label{eq:polarization}
\end{equation}
We compare with the result found in \cite{Ivanov2004}, obtained in
the case of the monochromatic wave, and see that in the short pulse
one reaches a slightly smaller value. However, if one collimates the
photon beam, one can achieve circular polarization to a degree close
to unity. The method presented here is particularly useful as
we require that $\xi$ is of the order of $1$ in such a way 
that the emission of high harmonics is suppressed. Moreover, at $\xi\sim 1$
the total probability of emission is of the order of $2\pi\alpha N$ \cite{Baier1998}
such that the obtained results are valid even for relatively long pulses as long as 
multiple photon emission is negligible. At the same time, this also implies that in the 
situations discussed above one cannot use the often used local constant field approximation, 
and the semiclassical method presented here is a simple method to obtain accurate values of 
the degree of polarization which is valid also for external fields of complex spacetime 
structure [see also Refs. \cite{PhysRevLett.113.040402,PhysRevA.91.042118,PhysRevA.95.032121} for an alternative applicable method].

\section{Polarization in a bent crystal}

Bent crystals can be used to steer an electron or positron beam along a circular arc as
investigated in e.g. \cite{PhysRevLett.114.074801,PhysRevAccelBeams.19.071001,PhysRevLett.119.024801,PhysRevLett.112.135503}. Also, the possibility of polarizing an electron/positron beam as in a storage ring through synchrotron radiation was discussed in e.g. \cite{Baryshevski1989}, where it was assumed that the crystal was bent close to the so-called Tsyganov critical radius which we will define as

\begin{equation}
R_{c}=\frac{\varepsilon d_{p}}{2U_{0}},\label{eq:critradius}
\end{equation}
where $d_{p}$ is the distance between two symmetry planes in the crystal 
and $U_{0}$ is the corresponding potential energy depth. This radius corresponds to the radius at which the strength of the force from the electric field between the planes, estimated as $2U_0/d_p$, can no longer provide the necessary centripetal force to sustain the circular motion.  Below we consider the motion of a positron between two (110) planes in Germanium such that $d_{p}=2.0$ {\AA}  and $U_{0}=35.73$ eV. According to the above discussion, the Tsyganov critical radius is roughly the smallest bending radius at which channeling is still possible in the crystal. In this case the radiation and polarization characteristics
are that of the constant magnetic field which produces the same bending
radius of the trajectory, and therefore the largest possible polarization
is given by $8/\left(5\sqrt{3}\right)$ \cite{sokolov1966synchrotron},
when $\chi\ll1$ \cite{Baier_1972}. Conversely, when the bending radius
becomes large, one must recover the case of the flat crystal, which
does not produce any beam polarization. With the presented approach
we demonstrate that one can predict the polarization properties for
any bending radius $R$, and not only for the extreme case close to the
critical radius. In an experiment the average polarization will depend
on the angular distribution of particles when entering the crystal. Thus,
we will only apply the approach in the case of a single particle starting with and angle of $0$ and a distance of
$u_0=0.083$ {\AA} from the plane (this value corresponds to the thermal vibrational amplitude of the nuclei in the crystal lattice). The maximum polarization that can be asymptotically obtained, $A$, is given by \cite{Baryshevski1989,RevModPhys.48.417}
\begin{figure}[t]
\includegraphics[width=1\columnwidth]{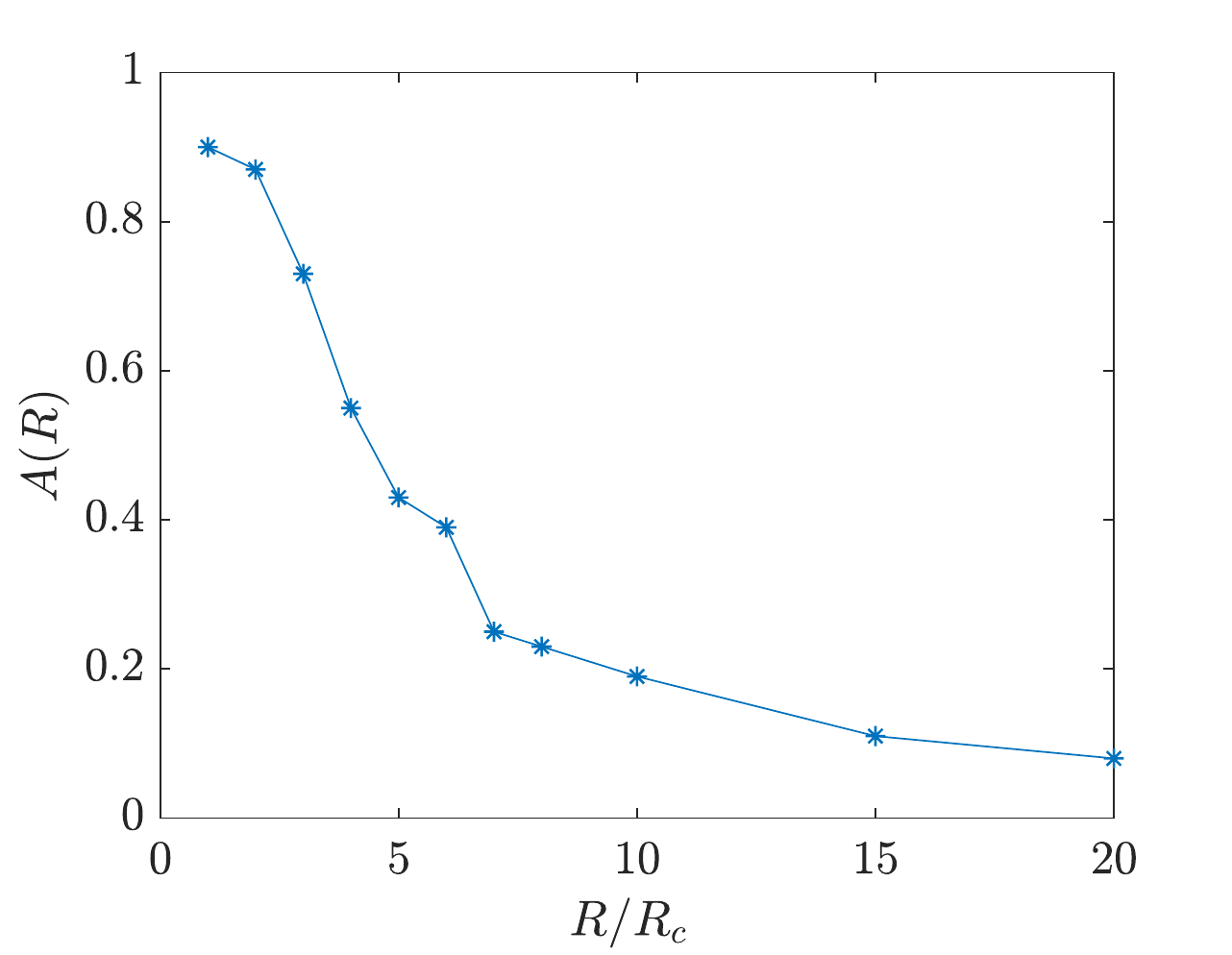}
\caption{The maximum possible transverse polarization that can be obtained
for a positron with the initial conditions mentioned in the text, depending
on the bending radius $R$ of the crystal in units of the Tsyganov critical radius $R_c$.\label{fig:fig4}}
\end{figure}

\begin{equation}
A=\frac{W_{\uparrow\downarrow}-W_{\downarrow\uparrow}}{W_{\uparrow\downarrow}+W_{\downarrow\uparrow}},\label{eq:maxpol}
\end{equation}
where $W_{fi}$ denotes the total transition rate from state $i$
to state $f$. The quantity $W_{fi}$ for different initial and final spin quantum numbers can be found from Eq. (\ref{eq:Baierfinal}) by integrating over angles and photon energies, and by summing over the photon polarization, using a finite piece of trajectory. This formula comes about if it is assumed that the positron has its energy replenished between each radiation emission, as is the case in a synchrotron. With crystals, this would require several thin crystals with accelerating structures in between.
We integrated the Lorentz force equation of motion using the electric field obtained from the continuum potential \cite{Lind65,Baier1998}, such that the electric field in the unbent crystal is along the $x$ direction. We then offset the plane along a circular arc in the $xz$ plane, which at the leading order in the small quantity $L/R$, where $L$ is the crystal length, means that the bending follows the curve $x=z^2/2R$. One may use this approximation as the total deflection angle $L/R$, is small in a realistic scenario. Due to symmetry, the electric field points along the radius of bending and using Gauss' law one can show that as long as the distance to the plane is much smaller than the bending radius $R$, the electric field component along the radius of bending is the same as the electric field in the unbent case evaluated at the same distance from the plane. The non-zero components of the electric field are then

\begin{equation}
E_{x}(x,y,z)=E_{\text{cont}}\left(x-\frac{z^{2}}{2R}\right),
\end{equation}

\begin{equation}
E_{z}(x,y,z)=-\frac{z}{R}E_{\text{cont}}\left(x-\frac{z^{2}}{2R}\right).
\end{equation}
Here, $E_{\text{cont}}(x-z^{2}/2R)$ is the electric field
obtained from the continuum potential, in the Doyle-Turner approximation
\cite{Doyle,avakian1982,Moller1995403}, which depends only on the
coordinate transverse to the planes (the $x$ coordinate in the considered case). We used a piece of trajectory
with roughly 10 periods of oscillation, which was adequate for convergence of the integrals. Moreover,
we have integrated over an angular region such that $\boldsymbol{v}_{\bot}(t)$ is contained in the region, with an additional angle of $10/\gamma$ in each direction. This turned out numerically
to be sufficient to cover all of the emitted radiation. In Fig.
(\ref{fig:fig4}) we show the result for a 50 GeV positron with the
mentioned initial conditions. It should be mentioned that this maximum
polarization is only achievable under the same circumstances as in
a storage ring, i.e. a short piece of crystal where radiation occurs,
and subsequently a replenishment of the lost energy so that the particles
have the nominal energy before entering a crystal again. It is seen that for a strong bending of the crystal, one approaches the
value in the constant field of $8/\left(5\sqrt{3}\right)$. While
we show only the example of a single trajectory, the method would
allow to study radiation reaction in a bent crystal where the effects
of polarization of the beam would be essential. We refer the reader to Refs. \cite{Wistisen2018experimental,Wistisen2019exp}
for recent experimental studies of radiation reaction in straight
crystals.

\section{Conclusion}

In conclusion we have presented a method to rewrite the semiclassical
formulas of Baier, Katkov and Strakhovenko, which facilities their numerical
implementation for arbitrary discrete particles quantum numbers. 
This then allows for the calculation of radiation emission
with arbitrary initial and final electron spins, and with arbitrary polarization
of the emitted photon when knowing only the classical trajectory of
the electron in the background field. In this way, one does not have to
know the Dirac wave function in the background field, which is typically an impossible task for realistic field configurations. 

First, we have compared the obtained formulas for a case where a solution of the Dirac equation is known, namely the plane-wave field, and find near perfect agreement between the two methods, corroborating the idea that the motion in a plane wave is intrinsically quasiclassical. As an example, we considered the case of the transfer of circular polarization of the radiation, when an electron beam head-on scatters on a short circularly polarized pulse, with the conclusion that the shortness
of the pulse implies a slightly lower degree of polarization as compared to the monochromatic-field case. However, much higher degrees of polarization are observed for the photons emitted approximately along the initial direction of propagation of the electrons, in agreement with angular momentum conservation. Finally, we considered the case of a bent crystal and showed how one can calculate the degree of polarization of the positron beam for an arbitrary bending radius of the crystal.

\section{Acknowledgements}

For T. N. Wistisen this work was supported by the Alexander von Humboldt-Stiftung.

\bibliography{biblio}

\end{document}